\documentstyle[aps,prl,epsfig,twocolumn]{revtex}
\title{Modified ``delta kick cooling'' for
studies of atomic tunneling}
\author{S.H. Myrskog, J.K. Fox,  H.S. Moon$^{*}$, H.A. Kim$^{*}$, J. B.
Kim$^{*}$ and A.M. Steinberg} 
\address{Department of Physics, University of Toronto\\
Toronto, ONT M5S 1A7 CANADA\\
$^{*}$ Permanent address: Dept. of Physics, Korea Nat.
Univ. of Education,\\
Chungbuk, Korea 363-791}
\begin{document}
\maketitle

(Submitted December 2, 1998)

\begin{abstract}
 We present progress towards a planned experiment on atomic tunneling of
ultra-cold Rb atoms. As a first step in this experiment we present a 
realization of an improved form of ``delta-kick cooling.''  By
application of a pulsed magnetic field, laser-cooled Rb atoms are further
cooled by a factor of 10 (in 1-D) over the temperature out of an optical
molasses. Temperatures below 700 nK are observed. The technique can be
used not only to cool without fundamental limit (but conserving
phase-space density), but also to
focus atoms, and as a spin-dependent probe. 
\end{abstract}

\section{Introduction}
Advances in laser cooling and atom optics have led both to the ability
to directly observe new physical effects (e.g., in connection with weakly
interacting Bose gases) and to develop new, potentially applicable
technologies (such as atom-wave gravimetry).  There has been a great
deal of interest in paving the way for future developments by studying
atom optical components such as mirrors, lenses, and beam-splitters
\cite{Aminoff=1993,Roach=1995,Pfau=1993,Prentiss=1994PRL,Ertmer=1997PRA}.
Often, these components
themselves rely on new and interesting
physical effects, which is one of the explanations for the self-sustaining
excitement in this field.

The long de Broglie wavelengths of ultracold atoms make them ideally
suited for studies of quantum mechanics, as evidenced for example by the
interference between two Bose condensates\cite{Andrews=1997}, or by tests
of quantum chaos using cold atoms\cite{Moore=1995}.  We plan to make use
of the wave nature of laser-cooled Rubidium atoms to study the tunneling
of atoms through optical dipole-force barriers; by focusing light into a
thin sheet, one can create strong repulsive potentials for atoms, with
spatial widths several times the optical wavelength.  By cooling atoms to
near or below the recoil temperature (where the atomic de Broglie
wavelength is equal to the optical wavelength), we can enter a regime
where there is significant tunneling probability.  Although tunneling has
been observed indirectly in several atom-optics
experiments\cite{Bharucha=1997PRA}, this would be to our knowledge the
first experiment where {\it spatially resolvable} tunneling is observed.  
That is, we plan to directly image the incident and transmitted atomic
clouds, using the internal degrees of freedom of the atoms to address
questions about the ``history'' of transmitted particles.  This unique
system should make it possible to answer long-controversial questions
related to tunneling and to quantum measurement
theory\cite{Tunnel,Steinberg=Microstructure}.  As a simple example, the
question of what you see if you attempt to image atoms while they are
within a forbidden region is already nontrivial\cite{Steinberg=Korea98}.

In addition to the intrinsic interest of atomic tunneling, focussed
dipole-force barriers may prove useful as coherent beam-splitters,
velocity-selective elements, and in related roles.  In this paper, we
present the status of our experimental project to observe atomic 
tunneling, including some simulations of planned velocity-selection
experiments.  As a first step towards this experiment, 
we have used time-dependent
magnetic forces to cool atoms in one dimension, in a scheme based loosely
on Ammann and Christensen's ``delta-kick cooling'' proposal\cite{Ammann}.
We have achieved temperatures below 700 nK, and much lower temperatures
are possible in principle.  The technique is also generalizable to
three dimensions.  We also see some interesting effects when these
magnetic kicks are applied to atoms from an optical molasses.  We
are still working towards a full understanding of these effects, but
they appear to contain the signature of spatial spin correlations
within the atom cloud.  We believe that in addition to the usefulness
of magnetic kicks for cooling, this Stern-Gerlach-like system will prove
to be a novel probe of the atomic spin in laser-cooled atom clouds.

As alluded to above, the practical observation of atomic tunneling will
require atoms with de Broglie wavelengths on the order of the width
of the tunnel barrier.  Our barrier is produced by using the dipole force
from a 500 mW laser focused into
a sheet of light which is about 10 $\mu$m wide and 2 mm tall.  The
atoms leaving our laser cooling and trapping system are at temperatures of
6 $\mu$K, implying a wavelength of about 0.2 $\mu$m, too short to observe
any significant tunneling through an optical-scale barrier. Therefore we
need to
further cool our atoms to sub-recoil temperatures.  Our first stage of
cooling following optical molasses is to perform an improved variant of
the ``delta kick
cooling'' proposed by Ammann and Christensen\cite{Ammann}, also similar
to an independent proposal by Chu et al. \cite{Chu=1986} . Following this
stage, we will have atoms with 1-D temperatures on the order of the recoil
temperature of Rb and de Broglie wavelengths on the order of the optical
wavelength of 780 nm.  We shall perform a final step of velocity
selection to select
only the least energetic atoms and separate them from the remaining
higher-energy atoms.  By using 1-D velocity selection we can retain about 7\% of
our atoms at 1/200th of the initial temperature.  Given the fact that we
are concerned with 1D temperatures this is more efficient than comparable
evaporative cooling in 3D. Moreover, the velocity selection will suppress
the thermal tails which could otherwise lead to ambiguity between
tunneling and thermal activation.

In the original proposal of ``delta-kick cooling'' Ammann and Christensen
suggested a form of cooling in which an atomic wavepacket prepared in a
minimum-uncertainty state within an optical lattice is first allowed to
freely expand for a short period to allow position to become correlated
with momentum.  Application of a position-dependent force from a standing
wave can then be used to reduce the mean momentum of the atoms.  Near the
bottom of the potential well the atoms essentially experience a harmonic
potential.  As the atoms expand beyond this harmonic region the
proposed cooling process breaks down.  Instead of a sinusoidal potential,
we consider a true harmonic potential, easily generated with magnetic
field coils.  This variant is not only easier to understand and to
implement; it is also immune to the original proposal's cooling limit.

Consider an ensemble of laser-cooled atoms initially confined within a
small region
of size $r_{i}$ with mean thermal velocity $v_{o}$.  At time $t=0$ the
trap is turned off and atoms are allowed to freely expand away from the
trap center.  After a time $t_{\rm f}$, long compared to $r_{i}/v_{o}$ the
atoms are located at positions essentially given by $x_{a}=v_{a}t_{\rm
f}$, where $v_{a}$ is the velocity of a particular atom.  Application of a
harmonic potential $U=\frac{1}{2}m\omega^{2}x^{2}$ for a short time will
apply an impulse to the atoms proportional to position $\Delta \vec{p}=-m
\omega^{2} \vec{x} t_{\rm k}$ where $t_{\rm k}$ is the duration of the
kick.  If this impulse is chosen to be equal to $-mx_{a}/t_{f} \approx
-mv_{a}$, all atoms will essentially be brought to rest.  The condition
for such an optimal kick is $t_{\rm f} t_{\rm k}= 1/\omega^{2}$.

In reality several factors place limits on the achievable
temperature.  
At time $t=0$ the atoms are not all located at the center of
the trap, but instead have some distribution of initial positions.
As the
ratio between the final size to initial size increases, the correlation
between position and momentum improves, allowing greater degrees of
cooling.
At any finite time, the correlations will be imperfect, preventing
the cooling from being optimal.
Practical considerations may limit the time for which the atoms can
be allowed to expand.

After free expansion, the typical atom is at a position on the order
of $r_{f} = \sqrt{r_{0}^{2} + (v_{0}t)^2}$, and has a velocity on the
order
of $v_{0}$.  Each individual velocity class has a spread in position given
by the
initial size $r_{0}$ of the cloud.  A transferred impulse proportional
to distance, and designed to cancel out the mean velocity of $v_{0}$
at a typical distance of $r_{f}$, will therefore transfer a {\it random}
velocity on the order of $v_{0} r_{0}/r_{f}$ to each velocity group of
atoms.
Indeed, a more careful phase-space treatment shows clearly that the
rms velocity of the kicked atoms decreases by a factor of $r_{0}/r_{f}$,
leading to a temperature reduction of $(r_{0}/r_{f})^2$.

 In other words, this technique is the moral equivalent of adiabatic
expansion.  At a practical level, however, it has the advantage that there
is no adiabatic criterion to meet.  To cool a cloud by a factor of $N$
would
require a time of $\sqrt{N} \tau_{0}$ where $\tau_{0}$ is a secular period
in the harmonic oscillator. A similar adiabatic expansion would need to be
accomplished slowly relative to the instantaneous secular period($N
\tau_{0}$ by the end of the expansion).

The cooling process can be readily understood in a phase-space picture.
Figure 1 shows
the evolution of an atomic cloud in phase space from the initial state at
the start of expansion, to after free expansion and finally, the
distribution after application of a kick.  We start with a phase-space
distribution characterized by the widths of the distribution in momentum
and position.
Free expansion stretches the distribution in position, but has no
effect on momentum.  The effect of a harmonic kick is to rotate the
distribution in phase space.
If the duration of the kick is chosen correctly, this rotates the
distribution back onto the position axis, thereby lowering the temperature.
Note that this does not require a true ``delta-kick'' in the sense of
a very short duration.
By Liouville's theorem, the area in phase space is
conserved both during free expansion and during the kick, 
yielding $v_{f} = v_{0} x_{0}/x_{f}$.
The
longer the cloud is allowed to undergo free expansion before the kick, the
narrower the final distribution is in momentum, resulting in lower
temperatures.

\begin{figure}
\centerline{\epsfig{file=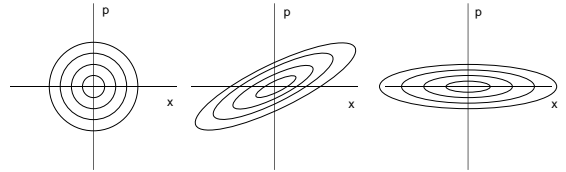,width=3in}}
\vspace{10pt}
\caption{Phase space diagrams.  a) is a distribution of the localized
cloud.  b) shows the cloud after a period of free expansion prior to a
kick.  c) A harmonic kick rotates the distribution onto the x-axis, lowering the
temperature of the cloud.}
\label{Figure 1}
\end{figure}

In addition to cooling via harmonic potentials, it is also possible to
cool atoms using a pulsed quadrupole potential.  Although this
technique cannot match the ideal harmonic kicks, it is substantially
easier to generate strong field gradients than higher-derivative
terms.  In 1D, a quadrupole field exerts a fixed impulse toward the center
of the potential.  If the atoms have been allowed to expand significantly,
most are moving away from the center, and a kick chosen to be equal to the
mean thermal velocity concentrates the atoms near v=0 (in a highly
non-thermal, nearly flat-topped distribution).  In 1D, the mean kinetic
energy may be reduced by up to a factor of 6 in this way.  Simulations for
a true 3D quadrupole potential show even greater cooling(predominantly
along the symmetry axis), due to the ``rounding'' of the potential energy
surface when
transverse excursions are taken into account.  The cooling is
predominantly along the quadrupole axis, but some cooling does occur in
the other directions as well.  

The data presented in this paper all concern one-dimensional cooling.
However, the harmonic kicks are easily generalizable to 3D by application
of successive kicks along the three Cartesian axes.  With quadrupole kicks,
perfect spherical symmetry cannot be achieved, but may be approximated
using a similar approach.

\section{Experiment}

Our MOT coils are located horizontally around the cuvette and define the
z-axis of the trap.  The coils are made of 200 turns of wire with a radius
of 4 cm and separated by 8 cm.  This is twice the separation  required for
a Helmholtz configuration, leading to a nonvanishing $d^{2}B/dz^{2}$.
With the
currents in opposing directions as used in a MOT or a quadrupole kick, the
coils can produce
gradients up to 180 G/cm given our present maximum current of 18 A.  These
same coils are also used for generation 
of a harmonic potential.  By reversing the direction of current  in one of
the coils we can achieve harmonic fields with a second derivative of about
60 G/cm$^{2}$, corresponding to a trap frequency of about 60 Hz. The coils
can be switched on and off in 200 $\mu$s, residual fields falling to 1\%
in 1 ms.

We start by cooling and trapping $^{85}$Rb atoms within our MOT using a
field gradient of 20 G/cm, and 40 mW of total power in trapping beams with
2 cm diameter. The trapping beam is detuned 10 MHz to the red of the D2
($F=3 \Rightarrow F=4$) transition at 780 nm.  We cool and trap about
$10^{8}$ atoms in our MOT with a diameter of 0.5 mm.  We further cool the
atoms to 6 $\mu$K in 1 millisecond of $\sigma^{+} \sigma^{-}$ optical
molasses detuned $34$ MHz to the red of the $F=3\Rightarrow F=4$
transition.

We then turn off all light and magnetic fields to allow the atoms to
undergo free expansion. After a time of 9 to 15 ms we apply a short (3 ms)
pulse of the quadrupole field.  The resulting force is directed  towards
the origin, mostly along the symmetry axis of the coils, and applies an
impulse to the atoms opposing their direction of motion.

The temperature of the resulting cloud is determined by time-of-flight
imaging.  A series of images is
taken as a function of time after the kick, and the temperature is
extracted from the expansion curve.

The existence of multiple spin levels in atoms released from an optical
molasses implies that
different atoms experience different magnetic potentials, making
distributions difficult to interpret. For this reason we select the atoms
in a doubly polarized state $F=3, m_{F}=3$ by capturing the atoms within a
weak magnetic trap which is unable to hold atoms with $m_{F}<3$ against
gravity.  We release the doubly polarized atoms from the magnetic trap
after 200 ms and allow the $m_{F}=3$ atoms to expand freely before
application
of the quadrupole magnetic field.

Figure 2 shows fluorescence images of the atoms under free expansion and
after application of a delta kick.  The atoms released from the magnetic
trap have a temperature of 7.5 $\mu$K with an rms velocity of 2.7 cm/s.  
The atoms are first allowed to undergo free expansion for a time of 11 ms
and then have a kick applied for 3 ms, imparting a change in velocity of
3 cm/s. After application of the kick, the temperature of the cloud along
the coil axis has decreased by a factor of about 6, from 7.5 $\mu$K down
to 1.2 $\mu$K.
Expansion curves for a cloud for several kick strengths are observed in
Figure 3.  
For a kick of 2.4 cm/s, near the original rms velocity of the atoms,
we observe optimal cooling.  Over 20 ms, essentially no expansion of
the atom cloud is seen, and fits indicate a temperature below 700 nK.
For stronger kicks, the atoms are impelled back to the center of the
cloud, where they come to a focus.
As the strength of the kick increases the atoms come to a smaller
focus (consequently heating the atom cloud), until the strongest kick
strength results in a cloud hotter than the cloud from the original
magnetic trap.

\begin{figure}
\centerline{\epsfig{file=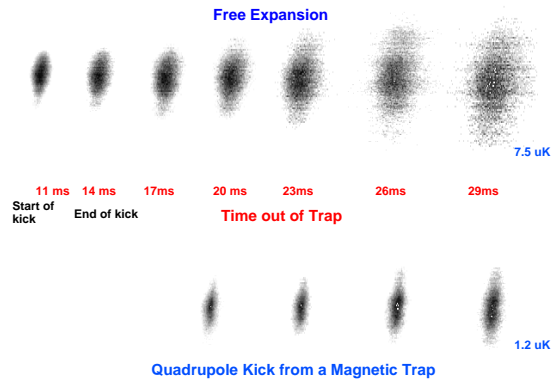,width=3in}}
\vspace{10pt}
\caption{A sample of the fluorescence images obtained.  The upper set of
images corresponds to atoms undergoing free expansion.  The lower set of
images corresponds to the atoms having undergone free expansion for 11ms
followed by a quadrupole kick for 3 ms, followed by further free expansion
for temperature measurement.  
The temperature of the atoms along the coil axis(horizontal) after a kick
has been reduced from 7.5 $\mu$K to 1.2 $\mu$K by application of the
kick.}
\label{Figure 2}
\end{figure}

\begin{figure}
\centerline{\epsfig{file=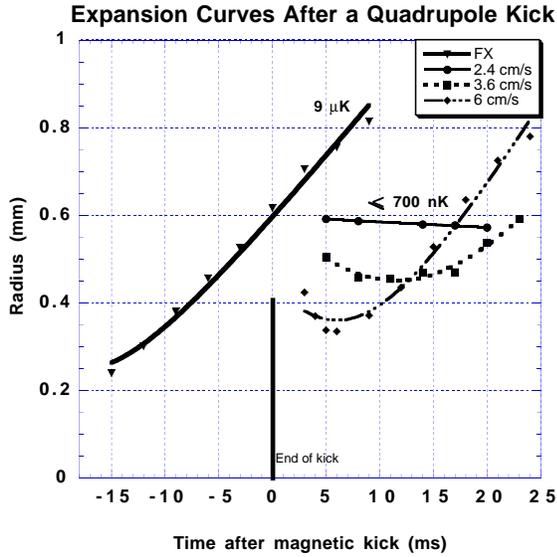,width=3in}}
\vspace{10pt}
\caption{The ballistic expansion of atom clouds is shown for free
expansion and after kicks of various kick strengths.
At high kick strengths the cloud can be brought to
a focus, with a tighter focus achieved as the strength of the kick is
increased.  For a nearly ideal kick, no expansion can be observed
over 20 ms; the temperature is lower than 700 nK.}
\label{Figure 3}
\end{figure}

\begin{figure}
\centerline{\epsfig{file=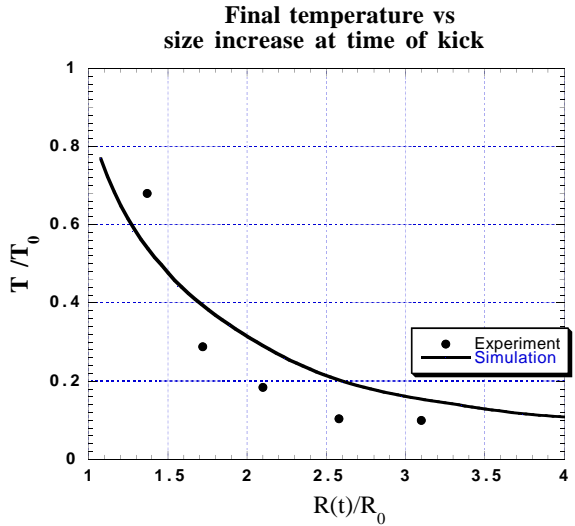,width=3in}}
\vspace{10pt}
\caption{A comparison of simulation and experimental data for the
ratio of final temperature to initial temperature as a function of the  
expansion ratio of the atomic cloud.  For short expansion
times the kick is unable to provide much cooling to the atomic cloud,
whereas at longer times (i.e. larger size ratio), cooling improves.
Experimental data shows cooling by a factor of 10, somewhat better
than originally predicted by the simulations, which neglect the effect
of gravity.}
\label{Figure 4}
\end{figure}

At short times such that the radius of the cloud has not increased by very
much, the positions and momenta of the atoms are not very correlated and
the kick does little to cool the atoms.  In Figure 4 we show results
(solid circles)
and simulations (smooth curve)
on the cooling ratio versus the ratio of final size to inital
size. 

\begin{figure}
\centerline{\epsfig{file=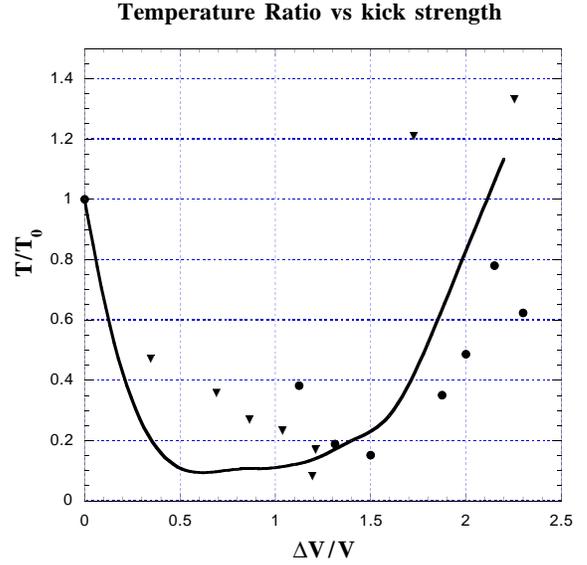,width=3in}}
\vspace{10pt}
\caption{Final temperature over initial temperature
as a function of the kicking strength.  The theoretical curve has a
minimum at a point where the change in velocity imparted by a kick is
equal to the rms velocity of the atomic cloud.  The experimental data
points from two different data sets (triangles and circles) yield
qualitative agreement, but a great deal of scatter due to sensitivity
to other initial conditions.  The minimum temperature appears to occur
for somewhat larger kick strengths than predicted by the simple
simulations; this can be understood by taking gravity into account.}
\label{Figure 5}
\end{figure}

The temperatures
obtained experimentally were lower than the initial temperature by as
much as a factor of 10, even at times when the simulations predicted
only a factor of 5.  
Additionally, Figure 5 suggests that optimal cooling occurs for a somewhat
stronger kick strength than originally expected theoretically.
Both these effects  can be qualitatively
understood by considering the effect of gravity.  When
the atoms are allowed to freely expand for a long time, they also begin to
fall under the influence of gravity.  If the cloud falls in the conical
potential by a distance larger than the spatial extent of the cloud, then
the horizontal potential seen by the atoms looks like a conic section,
i.e. a parabola.  Thus transverse cooling is accomplished by a harmonic
kick, much more efficient than the quadrupole, yet much stronger than the
true harmonic potential achievable by reversing the current in one of our
coils.

\begin{figure}
\centerline{\epsfig{file=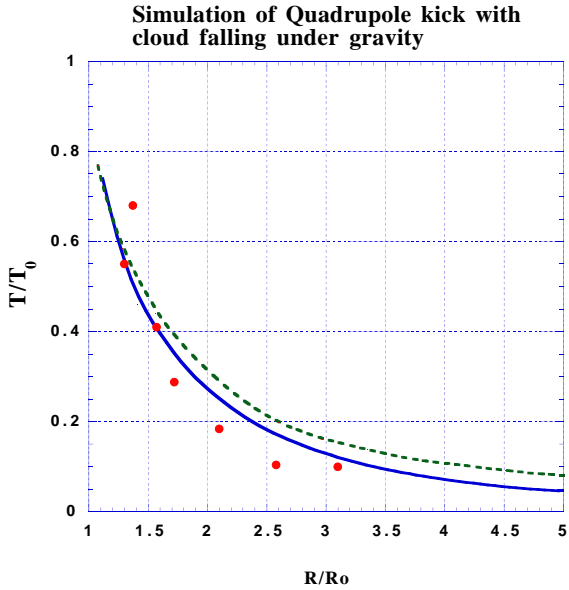,width=3in}}
\vspace{10pt}
\caption{The addition of gravity into our simulations demonstrates that
more cooling is possible than the quadrupole kick simulations suggest. 
The data from Fig. 5 are compared with a simulation neglecting gravity
(dashed curve) and one including gravity (solid curve).  
Clearly, the inclusion of gravity improves the cooling, as seen
experimentally.  The data are not perfectly modelled by this simulation
due to the strong dependence on details of the initial size and
temperature which may differ from point to point.}
\label{Figure 6}
\end{figure}

When we include gravity into our simulations, both the extra cooling and
increased potential strength requirements are observed.  Figure 6 shows
the achievable temperature with the inclusion of gravity. As the atoms
move down from the center of the potential, the strength of the horizontal
force
decreases.  To optimize cooling requires the use of a
stronger potential than expected for a 1D quadrupole.  Simulations
run with various parameters have shown cooling by as much as a factor
of 30.

Kicks have also been performed upon atoms coming out of an optical
molasses.  Since there are 7 different spin states in the F=3 ground state
of Rb, we expect the different states to undergo different amounts of
cooling or heating.  We therefore expect to see a multimodal distribution,
in which from the initial cloud radius, each spin component expands at a
different rate.  Figure 7 shows the results of a kick on an optical
molasses. The upper set of expansion images shown free expansion and the
lower set displays the atom cloud after a kick.  After the kick, the cloud
separates as expected into a bi-modal distribution consisting of a very
cold central stripe and a broader, hotter background.  The unusual feature
comes  in when we realize that the central stripe is
significantly smaller
than the size of the atom cloud at the time of kick as seen clearly in
Figure 8. We
believe this may
indicate a pre-existing correlation between position in the optical
molasses and spin state.  We are continuing to study these effects, for
which the Stern-Gerlach-like analysis possible using pulsed field
gradients appears to be a very promising new probe.

\begin{figure}
\centerline{\epsfig{file=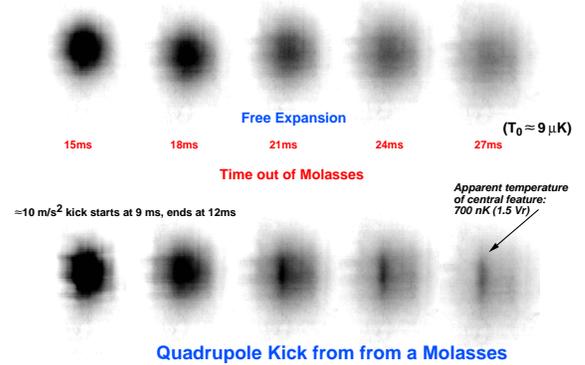,width=3in}}
\vspace{10pt}
\caption{ Free expansion and kicked atoms from a molasses.  Following
application of a kick a narrow stripe appears in the center of the cloud.
The temperature of the stripe is measured to be about 700 nK.  The width
of the stripe is smaller than the size of the cloud at the time of a kick,
possibly implying a pre-existing correlation between spins and position.}
\label{Figure 7}
\end{figure}

\begin{figure}
\centerline{\epsfig{file=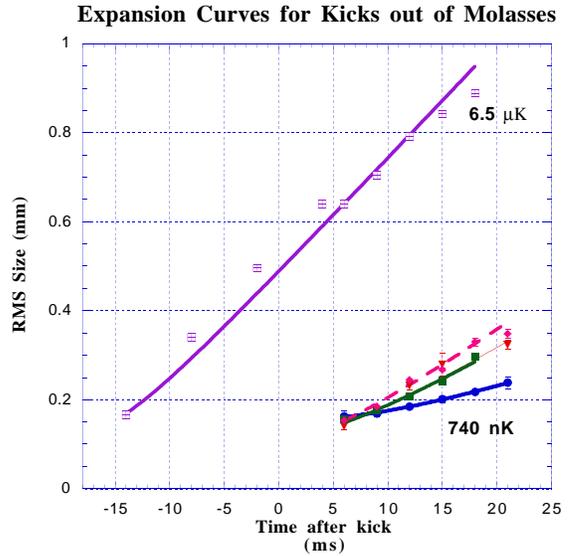,width=3in}}
\vspace{10pt}
\caption{  Radius vs time for a number of kick strengths.  The anomalously
small stripe size is clearly seen after a kick.  The lowest temperature
observed is about 700 nK.  }
\label{Figure 8}
\end{figure}

Through the use of these kicks we have achieved temperatures as low as 700
nK both on atoms released from a magnetic trap and on a
subset of the atoms released from an optical molasses.  Cooling ratios as
large as a factor of 10 have been observed.  The effect of gravity may be
used to enhance cooling, but also limits this technique for certain
applications.  Since the atoms begin to fall due to gravity, they are no
longer located at the origin of the potential, and cannot be recaptured
without heating.  The gain in cooling may still be useful for experiments
in which the atoms are allowed to fall for some distance, such as atomic
mirrors, lensing and atom deposition systems.  Along these lines 
Mar\'{e}chal et. al.\cite{Marechal=1998EPJD} have studied a system where
inhomogeneous magnetic potentials separate the spin components of a cloud
and provide cooling as the cloud falls a distance of about a meter.  
The addition of a 1D optical lattice or anti-gravity field will allow us
to reach greater ratios in the size of the atoms without having the atoms
fall from the center of the trap, to obtain cooling greater than a factor
of 10.

\section{Tunneling and Velocity Selection}

Having achieved recoil-velocity atoms we are moving on to study
properties of atoms while they are tunneling.  We shall use a line-focused
beam of intense light detuned far to the blue of the D2 line to create a
dipole-force potential for the atoms. Using a 500 mW laser at 770 nm
focused into a sheet 10 $\mu$m wide and 2 mm tall, we will be able to make
repulsive potentials with maxima as large as 50 $\mu$K, without
significant scattering rates.  The potential will be modulated using an
acousto-optic deflector allowing us to rapidly shift the focus of the
beam.  Since we may shift the focus of the beam more quickly than the
atomic center-of-mass motion can follow, the atoms see a time-averaged
potential.  This will allows us to make nearly arbitrary potential
profiles.

Following the application of a delta-kick to obtain sub-recoil
1-D velocities, our
atoms will have a thermal de Broglie wavelength of about 1 micron, still
too small to observe significant tunneling through a 10 $\mu$m
dipole-force barrier.  We will
therefore follow the delta kick with a velocity-selection phase.  We
will use
the same beam which is to form the tunnel barrier but dither the
focus of the laser quickly.  The atoms shall see a time-averaged potential  
many times larger than the de Broglie wavelength of the atoms,
thereby appearing as an essentialy classical barrier.
By sliding this
``classical'' barrier through our atomic cloud, it will be
possible to adiabatically sweep the
lowest energy
atoms away from the center of the magnetic trap, leaving the hot atoms
behind.  The coldest atoms will thus be in a spatially separated local
minimum from which tunneling may be observed when the width of the barrier
is decreased. Figure 9 shows
the results of quantum-mechanical simulations for atoms at an initial
temperature of 1.3 $\mu$K, trapped in a 5 G/cm field (300 $\mu$ K/cm).
We
superpose a 20 $\mu$m Gaussian beam with a peak height of 600 nK onto the
V-shaped potential.  This creates a potential minimum of 16 nK which is
swept through the magnetic trap at a rate of 0.5 mm/s.  The minimum
supports a quasi-bound state with energy 5 nK, and we see 7\% of the
atoms
transferred into this minimum.  Classically, we would expect the number of
atoms to be transferred equivalent to
$\sqrt{T_{f}/T_{i}}$.
In Figure 10 we observe steps in the probability of transfer as a function 
of the barrier height, indicating the number of quasi-bound states
supported.
\begin{figure}
\centerline{\epsfig{file=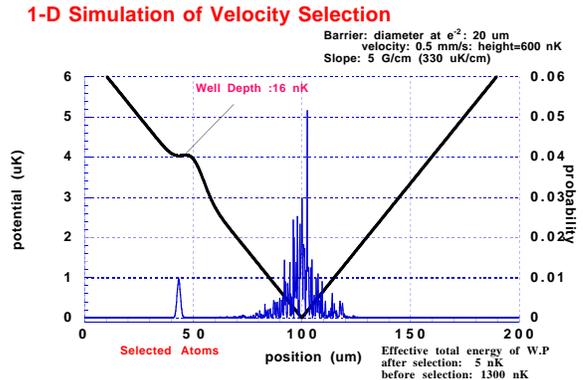,width=3in}}
\vspace{10pt}
\caption{   This quantum-mechanical simulation demonstrates what we expect to
achieve with our dipole-force velocity-selection.  Starting with an atom
cloud near 1 $\mu$K and sweeping an appropriately tuned 20 $\mu$m laser
beam through the atoms adiabatically, we will create a very small
auxiliary potential well capable of separating the coldest sub-sample 
of atoms from the cloud.  }
\label{Figure 9}
\end{figure}
\begin{figure}
\centerline{\epsfig{file=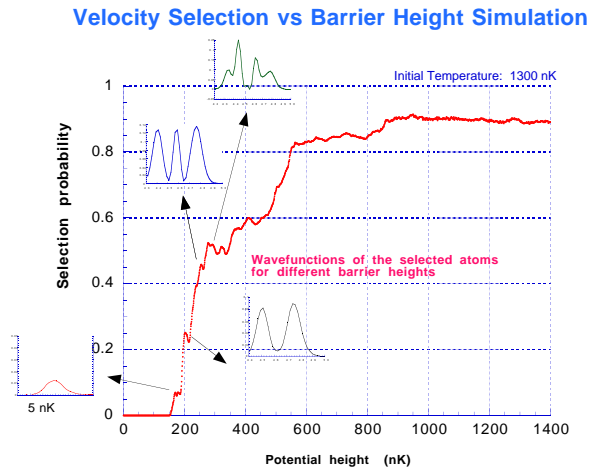,width=3in}}
\vspace{10pt}
\caption{   The probability of transfer exhibits steps as a
function of well depth, indicating the number of quasi-bound states
supported.  A well with a single bound state is seen to capture about 7\%
of the atoms, in a state with an energy of only 5 nK,
corresponding to a de Broglie wavelength of about 6 $\mu$m.  Such a state
will be an ideal source for our atom-tunneling experiments.}
\label{Figure 10}
\end{figure}

After velocity-selection we will narrow the beam to 10 $\mu$m to
allow the cold atoms to tunnel.  
The velocity-selected sub-sample of atoms will have a
de Broglie wavelength of 6 $\mu$m, leading to a significant tunneling
probability through a 10-micron barrier.  We expect rates of order 5\% per
secular period, causing the auxiliary trap to decay via tunneling on a
timescale of the order of 600 ms.

Progressing towards this tunneling experiment we have achieved
temperatures of 700 nK for atoms released both from
magnetic traps and optical molasses by application of a quadrupole
delta-kick.  Given longer expansion
times, this temperature should drop even further.
Once these cooling and velocity selection techniques are perfected, we
shall have a unique system in which to study tunneling.  By using optical
probes such as absorption, optical pumping and stimulated Raman
transitions, we will be able to
go beyond studies of simple wavepacket tunneling to investigate the
interactions of tunneling atoms while in the forbidden region.  Studies of
measurement of the tunneling time are planned as well as investigations
into quantum-mechanical nonlocality\cite{Steinberg=Foundations98}.


\begin{thebibliography}{99}
\bibliographystyle{unsrt}

\bibitem{Aminoff=1993}
C.G. Aminoff, A.M. Steane, P. Bouyer, P.Desboilles, J. Dalibard,C.
Cohen-Tannoudji, Phys. Rev. Lett. {\bf 71} 3083 (1993).

\bibitem{Roach=1995}
T.M. Roach, H. Abele, M.G. Boshier, H.L. Grossman, K.P. Zetie, E.A. Hinds,
Phys. Rev. Lett. {\bf 75} 629 (1995).

\bibitem{Pfau=1993}
T. Pfau, Ch. Kurtsiefer, C.S. Adams, J. Mlynek, Phys. Rev. Lett. {\bf 71}
3427 (1993).

\bibitem{Prentiss=1994PRL} 
J. Lawall, M. Prentiss, Phys. Rev. Lett. {\bf 72} 993 (1994)

\bibitem{Ertmer=1997PRA}
H. Hinderthur, A. Pautz, V. Rieger, F. Ruschewitz, J.L. Peng, K. Sengstock
and W. Ertmer, Phys. Rev. A {\bf 56} 2085 (1997)


\bibitem{Andrews=1997} M.R. Andrews, C.G. Townsend, H.J.. Miesner, D.S.
Durfee, D.M. Kurn and W. Ketterle, Science {\bf 275} 637 (1997)

\bibitem{Moore=1995}
F.L. Moore, J.C. Robinson, C.F. Bharucha, Bala Sundaram, and M.G.
Raizen, Phys. Rev. Lett. {\bf 75} 4598 (1995)

\bibitem{Bharucha=1997PRA}
C.F. Bharucha, K.W. Madison, P.R. Morrow, S.R. Wilkinson, Bala
Sundaram, and M.G. Raizen, Phys. Rev. A. {\bf 55} 857, (1997)



\bibitem{Tunnel}
M. B\"{u}ttiker and R. Landauer, Phys. Rev. Lett. {\bf 49} 1739 (1982)
; M. B\"{u}ttiker, Phys. Rev. B. {\bf 27} 6178 (1983)
; R. Landauer and Th. Martin, Rev. Mod. Phys. {\bf 66} 217 (1994)
; R. Chiao and A. Steinberg, Prog. Opt. {\bf XXXVII} 345 (1997)

\bibitem{Steinberg=Microstructure} 
A.M. Steinberg, Superlattices and Microstructures, {\bf 23} 823 (1998)

\bibitem{Steinberg=Korea98}
A.M. Steinberg, S. Myrskog, J.K. Fox, H.S. Moon, H.A. Kim, and J.B. Kim,
``On energy transfer by detection of a tunneling atom,'' in preparation


\bibitem{Ammann} H. Ammann, N. Christensen, PRL {\bf 78} 3072 1997

\bibitem{Chu=1986}
S.Chu, E. Bjorkholm,A. Ashkin,J.P. Gordon, L.W. Hollberg, Optics
Letters {\bf 11} 73 (1986).

\bibitem{Marechal=1998EPJD} \'{E}. Mar\'{e}chal, S. Guibal, J.-L.
Bossennec,M.-P. Gorza, R. Barb\'{e}, J.-C. Keller and O. Gorceix, Eur.
Phys. J. D {\bf 2} 195 (1998)

\bibitem{Steinberg=Foundations98}
A.M. Steinberg, Found. Phys. {\bf 28} 385 (1998)

\end{thebibliography}
 \end{document}